\documentclass[twocolumn,pra]{revtex4-1}
\usepackage{amssymb}
\usepackage{amsmath}
\usepackage{txfonts}
\usepackage{amsfonts}
\usepackage{graphicx}

\begin{document}

\title{Interference effect in the optomechanical stochastic resonance}
\author{Min Xie, Bixuan Fan, Xiaoli He, and Qingqing Chen}

\affiliation{College of Physics and Communication Electronics,
Jiangxi Normal University, Nanchang, 330022, China}

\begin{abstract}
In this paper, we study the stochastic resonance (SR) effect in an
optomechanical system driven by a strong coupling field and two weak signals
in both semiclassical and quantum frameworks. In the semiclassical description, the SR phenomena are found at
the cooperation of input signals and system noises. When two signals co-act
on our system, the interference effect between the optically induced SR and
the mechanically induced SR can be generated. In particular, a unique
beating effect, which makes the SR effect robust against the initial phase
difference of two signals, appears in the SR synchronization process with
unsynchronized signals. In addition, the quantum stochastic resonance effect
is numerically observed in the full quantum framework induced by pure
quantum fluctuations.
\end{abstract}

\maketitle

\section{Introduction}

In recent decades, stochastic resonance (SR) has attracted considerable
attention in various subjects \cite{rmp,rpp}, such as physics, chemistry,
biology and engineering science, for its intriguing and counterintuitive
behavior in nonlinear dynamical systems, whereby a subthreshold input signal
can be enhanced and optimized at an optimal noise level. SR was first
proposed in 1981 to explain the periodicity of the ice ages \cite%
{sr,climatic}, and since then it has been demonstrated theoretically and
experimentally in a variety of systems \cite{rmp,ufn}, and applied to weak
signal amplification \cite{amplification,amplification2} and detection \cite%
{detection,detection2}. A model most intensively investigated of SR is a
bistable system subject to a feeble periodic signal and noise \cite%
{bistable,bistable2}. It has been reported that SR can also occur in
monostable systems \cite{monostable,monostable2,monostable3} and multistable
systems \cite{multistable,multistable2,multistable3} as a consequence of
nonlinearities in those systems. In addition, the SR phenomenon has also
been extended into the quantum domain \cite{Coppersmith} and attracted
increasing attention, such as the spin-boson system \cite{Hanggi}, the
micromaser system \cite{Mantegna}, the dissipative anharmonic oscillator
\cite{Adamyan}, the quantum many-body system \cite{Plenio}, the Dicke model
\cite{Witthaut} and the Jaynes-Cummings model \cite{JC}.

It is known that nonlinearity is a key ingredient to induce the SR effect.
With the development of fabricating optomechanical devices, the radiation
pressure mediated optomechanical nonlinear coupling allows various nonlinear
effects, such as bistability \cite{bistability}, multistability \cite%
{multistable3,multistability,multistability2}, instability \cite%
{instability,instability2}, and chaos \cite{chaos,chaos2}. A standard
optomechanical system (OMS) consists of an optical cavity where one of the
end-mirrors oscillates and the radiation pressure on the moving mirror
creates a nonlinear interaction between the optical mode and the mechanical
mode. The study of SR in such a basic OMS may have great importance in
understanding the nature of the SR effect \cite{bistable2,OMS} and
application in weak-signal detection \cite{ws}.

In this paper we investigate the SR effect in an OMS subject to two weak
signals (optical and mechanical signals) in the semiclassical framework and
one weak mechanical signal in the quantum domain. In the semiclassical
regime, the SR effect activated by the white noise is studied in three
situations: a single signal, two synchronized signals, and two unsynchronized
signals. The results show that the system modulated by a single subthreshold
signal and a suitable noise can realize periodic interwell hopping
synchronized with the signal, which is the typical SR effect. Interestingly,
except for the conventional SR resonance peak in the signal-to-noise ratio
(SNR) curve, a stage of decrease appears for lower signal amplitudes at a
lower noise range due to the intrawell oscillation in a single well.

For the case of two signals, the system response can be interpreted as the
interference between the two signals induced SRs. The interference of SRs
usually occurs in the multistable systems \cite{multistable,multistable3}.
Here we present the interference of SRs occurring in a bistable system,
which is jointly induced by an optical channel and a mechanical channel. We
show that the constructive interference of two synchronized signals can
reduce signal amplitudes for inducing SR, and the beating-like effect can
appear when one signal is slightly detuned from the other. We find that the
SR effect is robust and insensitive to the initial phase difference of
signals as a result of beating.

In the quantum description, we explore the system stochastic dynamics
induced by a weak mechanical force and quantum fluctuations using the
quantum trajectory theory \cite{Milburn}. The results show that at zero
temperature the quantum stochastic resonance (QSR) can also be observed in
our bistable OMS and the system responses are synchronized to the external
signal under appropriate system parameters at the optimal signal frequency.

This paper is organized as follows. In Sec. II, we introduce our model, and
analyze the steady-state solutions as well as the stability. In Sec. III, we
show the SR effects induced by a single modulated signal and thermal noise
in the semiclassical framework, including the input-output synchronization
and the resonance peak in SNR curve. Then, the combined effect of two
signals in the SR process is discussed. In Sec. IV, we investigate the QSR
effects subject to a weak mechanical force and pure quantum noise at zero
temperature, and the residence time distribution and the system synchronous
responses to the signal are described. In Sec.V, our conclusions are
presented.

\section{Model}

We consider a standard OMS whereby the position of a mechanical oscillator
modulates the resonance frequency of an optical cavity, as shown in Fig. \ref%
{system}. A mechanical mode with resonance frequency $\omega _{m}$ and an
optical mode with frequency\ $\omega _{a}$\ are coupled through the
radiation pressure. The optical cavity is driven by a strong control field $%
E_{c}$\ with frequency $\omega _{c}$\ and a weak signal field $E_{s}$\ with
frequency $\omega _{s}$; a weak force $F_{s}$\ with frequency $\omega _{f}$ acts on the mechanical oscillator. In a rotating frame, the Hamiltonian
of the system reads ($\hbar =1$)
\begin{eqnarray}
\hat{H} &=&\frac{1}{2}\omega _{m}\left( \hat{x}^{2}+\hat{p}^{2}\right)
+\Delta \hat{a}^{\dag }\hat{a}+g\hat{a}^{\dag }\hat{a}\hat{x}  \notag \\
&&+E_{c}\left( \hat{a}^{\dag }+\hat{a}\right) +E_{s}\left( e^{-i\delta t}%
\hat{a}^{\dag }+e^{i\delta t}\hat{a}\right)  \notag \\
&&+F_{s}\cos \left( \omega _{f}t+\phi \right) \hat{x},  \label{hamiltonian}
\end{eqnarray}%
where $\hat{x}$ and $\hat{p}$ are the dimensionless position and momentum
operators of the mechanical mode; $\hat{a}^{\dag }\left( \hat{a}\right) $
are the creation (annihilation) operators of the optical mode; $g$ is the
optomechanical coupling; $\Delta =\omega _{a}-\omega _{c}$ is the detuning
between the optical mode and the strong coupling field; $\delta =\omega
_{s}-\omega _{c}$ is the frequency difference of the two external driving
fields $E_{s}$\ and $E_{c}$; whereas $\phi $ is the initial phase difference
of the two weak signals $F_{s}$\ and $E_{s}$.

\begin{figure}[th]
\centering\includegraphics[width=3.2in]{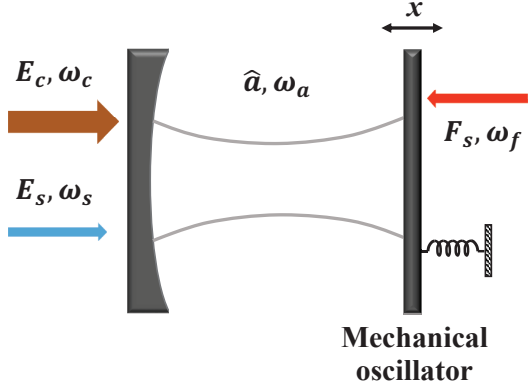}
\caption{Sketch of an optomechanical system. An optical resonator, driven by
a strong control field $E_{c}$ and a weak signal field $E_{s}$, is coupled
to the mechanical oscillator by radiation pressure. A weak force $F_{s}$
acts on the mechanical oscillator.}
\label{system}
\end{figure}

First, this paper deals with the system in the semiclassical description,
thus we neglect quantum fluctuations of optics and mechanics. By using the
Heisenberg equation of motion and phenomenologically adding thermal noise
and damping terms, we can obtain the mean value equations of motion for
classical system variables $\alpha =\left\langle \hat{a}\right\rangle $, $%
x=\left\langle \hat{x}\right\rangle $ and $p=\left\langle \hat{p}%
\right\rangle $:
\begin{eqnarray}
\dot{\alpha} &=&-\left( i\Delta +\kappa \right) \alpha -igx\alpha
-iE_{c}-iE_{s}e^{-i\delta t},  \label{me1} \\
\dot{\alpha}^{\ast } &=&\left( i\Delta -\kappa \right) \alpha ^{\ast
}+igx\alpha ^{\ast }+iE_{c}+iE_{s}e^{i\delta t},  \label{me2} \\
\dot{p} &=&-\gamma _{m}p-\omega _{m}x-g\left\vert \alpha \right\vert
^{2}-F_{s}\cos \left( \omega _{f}t+\phi \right) +\xi _{m}\left( t\right) ,
\label{me3} \\
\dot{x} &=&\omega _{m}p,  \label{me4}
\end{eqnarray}%
where $2\gamma _{m}$ is the mechanical damping rate, $2\kappa $ is the
optical decay rate and $\xi _{m}$ is the stochastic noise acting on the
mechanical oscillator. For a high mechanical quality factor $Q=\omega
_{m}/\gamma _{m}>>1$, $\xi _{m}$ is the stochastic white noise and it obeys
the $\delta$-correlation $\left\langle \xi _{m}\left( t\right) \xi _{m}\left(
t^{^{\prime }}\right) \right\rangle =2D\delta \left( t-t^{^{\prime }}\right)
$ with the strength of noise $D\simeq \frac{\gamma _{m}}{2}\left( 2\bar{n}%
+1\right) $, where $\bar{n}=\left[ \exp \left( \hbar \omega
_{m}/k_{B}T\right) -1\right] ^{-1}$\ \ is the mean thermal excitation number%
\cite{Kac,milburn2}. The thermal optical noise can be ignored at low
temperatures as the thermal occupation of the optical mode is far below one.

By setting the time derivatives in Eqs. (\ref{me1})-(\ref{me4}) to zeros, we
can obtain the steady-state equation for mechanical position $x_{s}$,
\begin{equation}
g^{2}x_{s}^{3}+2g\Delta x_{s}^{2}+\left( \Delta ^{2}+\kappa ^{2}\right)
x_{s}+\frac{g\omega _{m}E_{c}^{2}}{\left( \omega _{m}^{2}+\gamma
_{m}^{2}\right) }=0,  \label{toe}
\end{equation}%
which is a cubic equation of $x_{s}$. As a consequence, three solutions of $%
x_{s}$ may exist in a certain range of system parameters, providing the
possibility for bistability.

Following the linear stability analysis, we can rewrite the system operators
as a sum of their steady-state values and zero-mean fluctuations, i.e., $%
\hat{y}\rightarrow y_{s}+\hat{y}$, and obtain the linearized equations of
motion by ignoring high-order terms of fluctuations:
\begin{equation}
\dot{y}=J\hat{y}+\xi ,  \label{le}
\end{equation}%
where $\hat{y}=\left[ \hat{a},\hat{a}^{\dag },\hat{p},\hat{x}\right] ^{T}$, $%
\xi =\left[ -iE_{s}e^{-i\delta t},iE_{s}e^{i\delta t},-F_{s}\cos \left(
\omega _{f}t+\phi \right) +\xi _{m}\left( t\right) ,0\right] ^{T}$, and the
Jacobian matrix $J$ is given by
\begin{equation}
J=\left[
\begin{array}{cccc}
-i\left( \Delta +gx_{s}\right) -\kappa & 0 & 0 & -ig\alpha _{s} \\
0 & i\left( \Delta +gx_{s}\right) -\kappa & 0 & ig\alpha _{s}^{\ast } \\
-g\alpha _{s}^{\ast } & -g\alpha _{s} & -\gamma _{m} & -\omega _{m} \\
0 & 0 & \omega _{m} & 0%
\end{array}%
\right] .  \label{J}
\end{equation}%
The criterion of a stable solution is that the real parts of all eigenvalues
of the Jacobian matrix $J$ are negative.
\begin{figure}[th]
\centering
\includegraphics[width=3.2in]{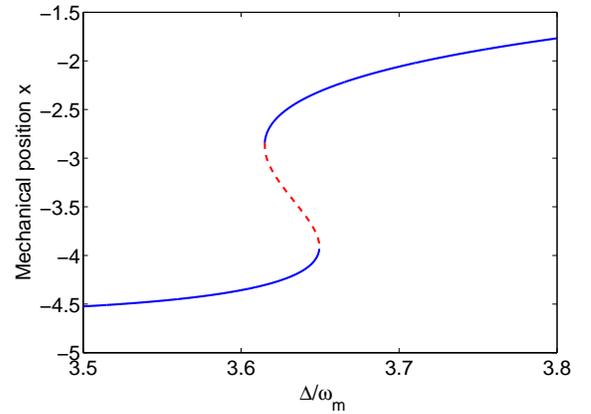}
\caption{The system stability diagram. Blue stands for the stable branches,
and red stands for the unstable branch. The parameters are $\protect\kappa =2.0%
\protect\omega _{m}$, $\protect\gamma _{m}=2\times 10^{-4}\protect\omega %
_{m} $, $E_{c}=5.05\protect\omega _{m}$ and $g=0.72\protect\omega _{m}$.}
\label{bistable}
\end{figure}

The mechanical position $x$ versus the detuning $\Delta $\ is illustrated in
Fig. \ref{bistable}. It is clear that the system exhibits mechanical bistability and the
mechanical position undergoes a transition from a single solution to three
solutions. In the three-solution region, the upper and lower branches correspond to two stable solutions and the middle branch is unstable. In the following, we are
interested in system dynamics in this bistable region.

To investigate the dynamics of the mechanical mode, we approximately derive
the equation of motion merely for the mechanical mode under $\kappa >>\gamma
_{m},g$, where the dynamics of the optical mode is much faster than that of
the mechanical mode and $\dot{\alpha}$ ($\dot{\alpha}^{\ast }$)\ can be set
to zero to solve the stable value of $\alpha $ ($\alpha ^{\ast }$). In this
case, the equation of motion for the mechanical mode can be simplified to
\begin{eqnarray}
\ddot{x}+\gamma _{m}\dot{x} &=&-\omega _{m}^{2}x-\omega _{m}\left[ g\frac{%
\left\vert E_{c}+E_{s}e^{-i\delta t}\right\vert ^{2}}{\left( \Delta
+gx\right) ^{2}+\kappa ^{2}}\right.  \notag \\
&&\left. +F_{s}\cos \left( \omega _{f}t+\phi \right) -\xi _{m}\left(
t\right) \right] .  \label{dynamics}
\end{eqnarray}%
From this, we can obtain the effective potential function for the position
of the mechanical oscillator in the absence of noise,
\begin{eqnarray}
U(x) &=&\frac{1}{2}\omega _{m}^{2}x^{2}+\omega _{m}F_{s}\cos \left( \omega
_{f}t+\phi \right) x  \notag  \label{potential} \\
&&+\frac{\omega _{m}\left\vert E_{c}+E_{s}e^{-i\delta t}\right\vert ^{2}}{%
\kappa }\arctan \frac{\Delta +gx}{\kappa }.
\end{eqnarray}

\begin{figure}[th]
\centering\includegraphics[width=3.2in]{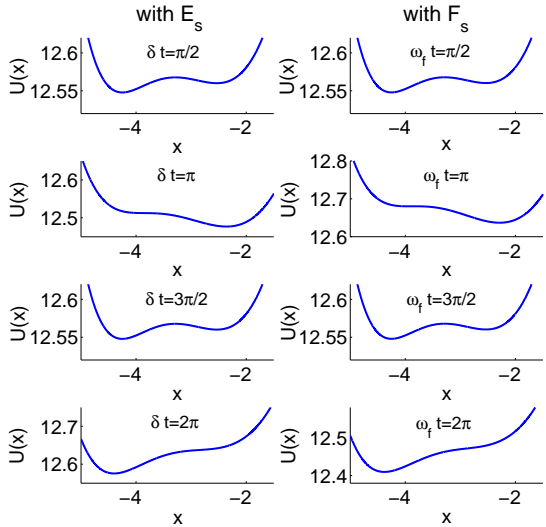}
\caption{Effective potential function of the mechanical position in one
period with the optical signal $E_{s}=0.021\protect\omega _{m}$ (left panel)
or the mechanical force $F_{s}=0.032\protect\omega _{m}$ (right panel). The
detuning $\Delta =3.626\protect\omega _{m}$, the frequency difference $%
\protect\delta =0.0006\times 2\protect\pi \protect\omega _{m}$\ and the
mechanical signal frequency $\protect\omega _{f}=0.0006\times 2\protect\pi
\protect\omega _{m}$, the initial phase difference $\protect\phi =0$. The other
parameters are the same as those in Fig. \protect\ref{bistable}.}
\label{f-potential}
\end{figure}

As shown in Fig. \ref{f-potential}, the effective potential varies
periodically in the presence of the optical signal $E_{s}=0.021\omega_m$
(left panel) or the mechanical force $F_{s}=0.032\omega_m$ (right panel).
Assuming that the two signals are synchronous over time, i.e., $\phi =0$, we
can see that both signals give periodic modulation on the potential function
and the modulations from the two signals are synchronized.

\section{Stochastic resonance in the semiclassical framework}

In this section, we present our main results in the semiclassical
description: SR phenomena of our system under different thermal noise and
driving signals. In all simulations, we assume that the mechanical
oscillator is initially located at the original coordinate, i.e., $x(t=0)=0$%
. To observe noise-induced system responses, the signals are chosen to be
below thresholds. That means the mechanical oscillator can not cross the
potential barriers only driven by the signals.

\begin{figure}[tbp]
\centering\includegraphics[width=3.2in]{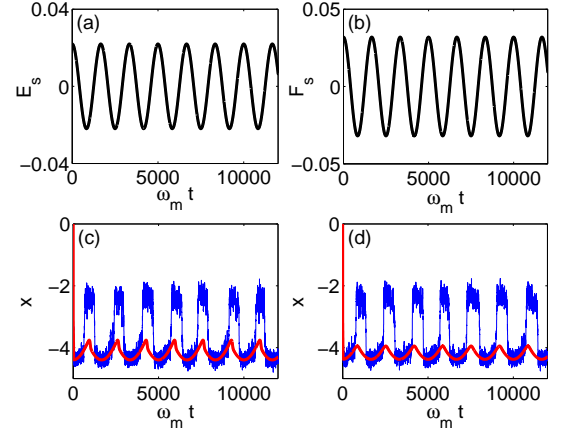}
\caption{Stochastic resonance with single modulation signal $E_{s}$ (left
panel) or $F_{s}$ (right panel) in the semiclassical description. (a) and
(b) are time evolutions of the signals; (c) and (d) present the steady-state
position of the mechanical mode without mechanical thermal noise $D=0$ (red
solid curves) and with $D=0.003\omega_{m}$ (blue curves). The other parameters are the
same as those in Fig. \protect\ref{f-potential}.}
\label{ss}
\end{figure}

Figure \ref{ss} presents the system dynamics due to a single modulation signal
$E_{s}$ (left panel) and $F_{s}$ (right panel), respectively, for the mechanical position $%
x $. It is clear that, in the absence of noise, the signals are
too weak to drive the mechanical oscillator from one potential well into the
other, and they can only drive small-amplitude oscillations within a single
well as shown in the red curves in Figs. \ref{ss}(c) and \ref{ss}(d). By
adding a certain amount of thermal noise $D=0.003\omega_{m}$ to the system, the
noise-assisted hopping between the double potential wells can be observed,
and the hopping is exactly synchronized with signal frequencies. This is a
typical signature of the SR effect.

Except for the input-output synchronization, a resonance peak in the
relation of the SNR versus noise is another signature of SR. We now analyze this
feature with the mechanical signal $F_{s}$ only under different intensities.
Here we adopt the standard definition of the SNR: the signal in the power
spectrum divided by the noise background at the driving signal frequency,
i.e., $SNR=P_{s}/P_{n}$ \cite{rpp}.

\begin{figure}[tbp]
\centering\includegraphics[width=3.2in]{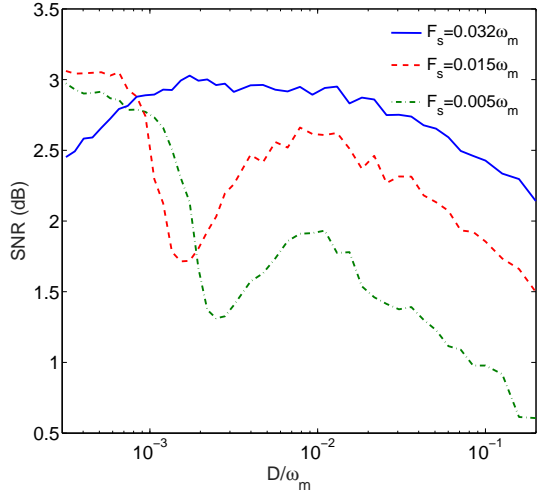}
\caption{The SNR in the decibel unit versus the noise intensity $D$ for different
amplitudes of the mechanical signal $F_{s}$. The other parameters are the
same as those in Fig. \protect\ref{ss} except for $E_{s}=0$.}
\label{snr}
\end{figure}
In Fig. \ref{snr}, the SNR of the mechanical signal versus the noise
strength is plotted for three different amplitudes of the signal. For $%
F_{s}=0.032\omega _{m}$, we can see a clear resonance peak in the SNR curve
and the trend is quite standard. It is interesting to note that, for lower
signal amplitudes ($F_{s}=0.015\omega _{m}$ and $F_{s}=0.005\omega _{m}$),
the SNR first experiences a stage of decrease other than the main SR
resonance peak as the noise increases. In this decreased stage the mechanical
oscillator is actually performing intrawell oscillation around the localized
potential minima since the noise is too low to induce the interwell
transition. As the noise intensity $D$ increases, the noise-assisted
interwell hopping occurs, and the main interwell SR peak appears.

\begin{figure}[th]
\centering\includegraphics[width=3.2in]{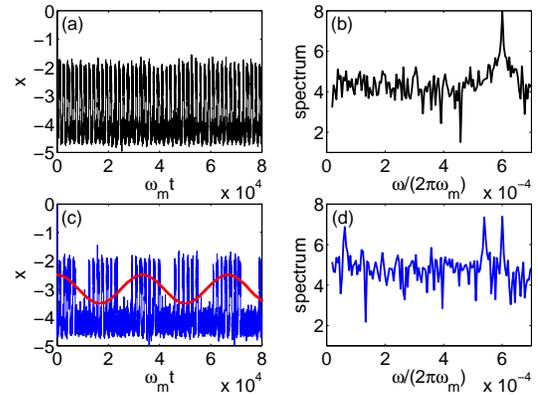}
\caption{The SR synchronization phenomena in the presence of two signals for
(a) $\protect\delta =\protect\omega _{f}$\ and (c) $\protect\delta -\protect%
\omega _{f}=\protect\delta /10$. (b) and (d) are the corresponding Fourier
spectrums. The red solid curve in (c) is the function of $0.5\cos \left(
\Delta \protect\omega t/2\right) -3.0$. The other parameters are the same as
those in Fig. \protect\ref{ss}\ except $E_{s}=0.012\protect\omega _{m}$ and $%
F_{s}=0.015\protect\omega _{m}$.}
\label{ts}
\end{figure}

The system dynamics driven by a single mechanical or optical signal has been
analyzed above. Now we turn our attention to the situation of two signals
simultaneously acting on the system. Figure \ref{ts} shows the mechanical
response of the system in the presence of two signals, and shows how their
frequency difference affects the SR effect. First we consider the situation
that two signals have the same modulation frequency and phases. With the
matched modulation frequencies ($\delta =\omega _{f}$) and appropriate
noise, the periodic hopping between two stable states can be observed in
Fig. \ref{ts}(a). The corresponding spectrum on the logarithmic scale is
shown in Fig. \ref{ts}(b), where a single peak is centered at the signal
frequency of $0.0006\times 2\pi \omega _{m}$. It can be easily explained as
the constructive interference caused by two synchronized signals. In
addition, compared to the single signal case, the amplitudes of two signals
required for SR to occur are substantially decreased, which is beneficial to
the detection of weak signals in experiments. And the system parameters we
have used are feasible for current experiment conditions \cite{Aspelmeyer}.

When the modulation frequency difference $\Delta \omega =\delta -\omega _{f}$
is a small but nonzero value, some interesting phenomena take place, as
shown in Figs. \ref{ts}(c) and \ref{ts}(d). Here we choose $\Delta \omega
=\delta /10$. The interplay of the two signals results in a complicated
beatinglike phenomenon where a slow modulation envelope and fast interwell
and intrawell hoppings coexist. It is clear that the period of the slow
envelope matches well with the curve $\mathrm{{cos}(\Delta \omega t/2)}$,
which is consistent with the theory of the beating signal. We can see that, in
the regions with large modulation amplitude, the interference between the
signal-induced responses is constructive and the system experiences
periodic interwell transition. In contrast, for the regions with low
modulation amplitude, their interference is destructive, and hence the
mechanical oscillator can not cross the potential barrier and it oscillates
inside a single well. Correspondingly, there are
multiple peaks in the frequency domain for $\Delta \omega \neq 0$ [see Fig.%
\ref{ts}(d)]. Two main signal peaks at the input signal frequencies $\delta $
and $\omega _{f}$ and a difference frequency signal peak at $\delta /10$\
can be seen. Furthermore, the main resonance peak at $\delta $ in Fig.\ref%
{ts}(d) is lower than that in Fig. \ref{ts}(b). It is not a surprising result
since the interference of two signals is the strongest when they are exactly
synchronized, as discussed in Ref. \cite{multistable3}.

\begin{figure}[th]
\centering\includegraphics[width=3.2in]{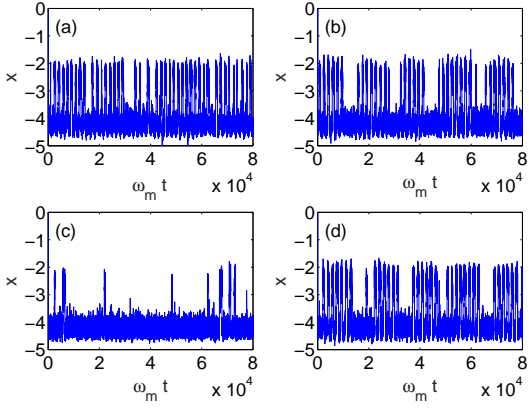}
\caption{The mechanical responses for several values of initial phase
difference $\protect\phi $ without (left panel) or with (right panel)
frequency difference. (a) $\protect\phi =\protect\pi /2,\Delta \protect%
\omega =0$; (b) $\protect\phi =\protect\pi /2,\Delta \protect\omega =\protect%
\delta /10$; (c) $\protect\phi =\protect\pi ,\Delta \protect\omega =0$; (d) $%
\protect\phi =\protect\pi ,\Delta \protect\omega =\protect\delta /10$. The other
parameters are the same as those in Fig. \protect\ref{ts}.}
\label{f-phase}
\end{figure}

Finally, we analyze the influence of the initial phase difference $\phi $\
of the optical and mechanical signals on the SR phenomenon. From Fig.\ref%
{f-potential}, we know that when two signals are initially synchronized,
i.e, $\phi =0$, their modulations on the potential function have the same
pace and therefore they cause the best constructive interference in the SR
phenomenon. For the other initial phases, their influences will be partly or
fully canceled. To confirm this effect, we plot the mechanical responses
for $\Delta \omega =0$ with $\phi =\pi /2$ and $\phi =\pi $ in Figs. \ref%
{f-phase}(a) and \ref{f-phase}(c). It is obvious that the SR effect
diminishes to vanish as $\phi $\ varies from $0$ to $\pi $. However, the situation is dramatically different when the
frequency difference of two signals is nonzero, i.e., $\Delta \omega =\delta
/10$. As shown in Figs. \ref{f-phase}(b) and \ref{f-phase}(d), the
beating-like phenomenon always exists for different initial phases, and the
input-output synchronization remains as good as that in Fig. \ref{ts}(c).
Therefore, the beatinglike phenomenon can make the synchronization behavior
and the SR effect more robust to the initial phase fluctuations.

\section{Quantum stochastic resonance in the full quantum framework}

In the preceding sections, we have studied the SR phenomena in the
semiclassical framework. Now we turn to investigate the QSR effect induced by pure quantum fluctuations using the
quantum trajectory method \cite{Milburn} at zero temperature. For a single
trajectory, the system dynamic conditioned on noisy homodyne detection can
be described by the stochastic master equation ($\hbar =1$),
\begin{eqnarray}
d\rho \left( t\right)  &=&dt\left\{ i\left[ \rho \left( t\right) ,\hat{H}%
\right] +\mathcal{D}\left[ \sqrt{2\kappa }\hat{a}\right] \rho \left(
t\right) +\mathcal{D}\left[ \sqrt{2\gamma _{m}}\hat{b}\right] \rho \left(
t\right) \right\}   \notag \\
&&+dW\left( t\right) \mathcal{H}\left[ \sqrt{2\kappa }\hat{a}\right] \rho
\left( t\right) ,  \label{sme}
\end{eqnarray}%
where $\rho \left( t\right) $\ is the density operator, $\hat{H}$\ is the
Hamiltonian of the OMS given in Eq.(\ref{hamiltonian}) and $dW$ is the
Wiener increment, satisfying $\left\langle dW\right\rangle =0$\ and $%
\left\langle \left( dW\right) ^{2}\right\rangle =dt$. The superoperators $%
\mathcal{D}$ and $\mathcal{H}$ are defined as
\begin{eqnarray}
\mathcal{D}\left[ \hat{A}\right] \rho  &=&\frac{1}{2}\left( 2\hat{A}\rho
\hat{A}^{\dag }-\hat{A}^{\dag }\hat{A}\rho -\rho \hat{A}^{\dag }\hat{A}%
\right) ,  \label{so1} \\
\mathcal{H}\left[ \hat{A}\right] \rho  &=&\hat{A}\rho +\rho \hat{A}^{\dag }-%
\text{Tr}\left[ \hat{A}\rho +\rho \hat{A}^{\dag }\right] \rho .  \label{so2}
\end{eqnarray}%
The corresponding homodyne detection current is given by
\begin{equation}
I\left( t\right) =\sqrt{2\kappa }\left\langle \hat{a}+\hat{a}^{\dag
}\right\rangle +dW\left( t\right) /dt.  \label{homodyne}
\end{equation}

\begin{figure}[th]
\centering\includegraphics[width=3.2in]{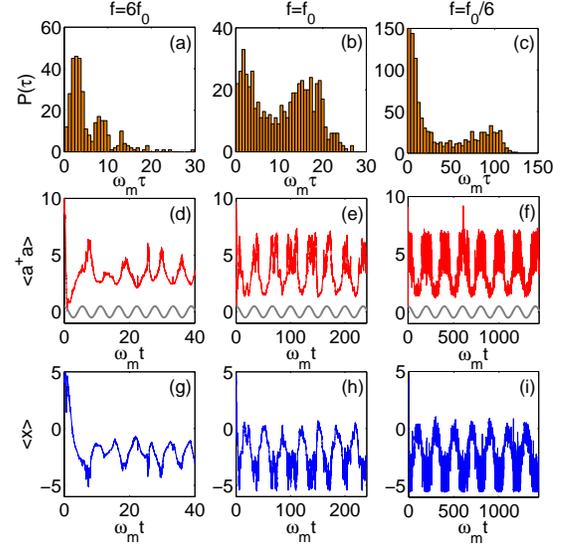}
\caption{The system responses subjected to a weak force and the quantum
noise for three different modulation frequencies. (a-c) Histograms for the
residence time distributions; (d-f) the average photon numbers. The gray
curve is the function of $0.5\cos \left( \protect\omega _{f}t\right) $,
characterized the period of the weak force; (g-i) the mechanical positions.
The parameters are $E_{c}=5.15\protect\omega _{m},$ $E_{s}=0,$ $F_{s}=%
\protect\omega _{m},$ $\protect\gamma _{m}=0.3\protect\omega _{m}$ and $%
f_{0}=0.03$. The other parameters are the same as those in Fig. \protect\ref{ss}%
. }
\label{f-qsr}
\end{figure}

For convenience, we only investigate the QSR effect induced by a
subthreshold weak force $F_{s}$ and the quantum noise. In Figs. \ref{f-qsr}
(a)-(c), we present the residence time distributions subject to the periodic
weak force $F_{s}\cos \left( \omega _{f}t\right) $\ for three different
modulation frequencies $f=6f_{0},f_{0},f_{0}/6$, where $\omega _{f}=f\times
2\pi \omega _{m}$\ and $f_{0}=0.03$. The results show that the resonance can
be achieved under appropriate parameters by varying the modulation
frequency, distinguished by a separate peak of the distribution [see Fig. %
\ref{f-qsr} (b)].

In Figs. \ref{f-qsr} (d)-(i), we show a few representative trajectories of
the photon number ($\left\langle \hat{a}^{\dag }\hat{a}\right\rangle $) and
the mechanical position ($\left\langle x\right\rangle $). The system
responses are synchronized to the signal best at the optimal modulation
frequency $f_{0}$; a higher frequency $6f_{0}$ or a lower frequency $%
f_{0}/6$ leads to the destruction of synchronization, similar to the
responses of the SR effect activated by thermal noise. However, due to
different noise levels and system parameters, it requires a different time
scale or frequency scale of the signal to satisfy the QSR matching condition.

\section{Conclusion}

To summarize, we have investigated noise induced synchronization to external
signals in a bistable optomechanical system in the semiclassical and quantum
frameworks. Either a single optical signal or a single mechanical signal can
induce the SR effect in our system. When the two external signals act on the
system jointly, we can observe an interference of SRs, which leads to the
beatinglike phenomenon depending on the frequency difference between
signals. In addition, due to the beating-like effect, the input-output
synchronization is more robust against the initial phase difference of two
signals. Our results reveal that the optical pathway can be utilized to
control the mechanical SR effect and detect the weak mechanical signal in a
basic optomechanical system. Besides, we have numerically demonstrated the
QSR effect induced by a weak force and pure quantum noise using the quantum
trajectory method. The QSR effect, similar to the SR effect induced by the white
noise, can be obtained under different system parameters at the optimal
modulation frequency.

\section*{Acknowledgments}

The authors would like to thank Dr. Z. Duan for valuable suggestions.
We gratefully acknowledge financial support from the National Natural
Science Foundation of China under Grants No. 11504145 and No. 11664014 and
from the Natural Science Foundation of Jiangxi Province under Grants No.
20161BAB211013 and No. 20161BAB201023.

\end{document}